\documentclass[prb,twocolumn,showpacs]{revtex4}
\usepackage{amsmath,amsfonts,amsthm,graphicx}
\usepackage{graphicx}
\usepackage{graphics}
\usepackage{dcolumn}
\usepackage{bm}

\newcommand{\beq}{\begin{equation}}
\newcommand{\eeq}{\end{equation}}
\newcommand{\beqnar}{\begin{eqnarray}}
\newcommand{\eeqnar}{\end{eqnarray}}
\newcommand{\bfig}{\begin{figure}}
\newcommand{\efig}{\end{figure}}

\begin{document}
\title{Spin polarization and magnetoresistance through a ferromagnetic barrier in bilayer graphene}

\author{Hosein Cheraghchi}
\email{cheraghchi@du.ac.ir}  \author{Fatemeh Adinehvand}
\affiliation{School of Physics, Damghan University, 6715- 364,
Damghan, Iran}
\date{\today}
\newbox\absbox

\begin{abstract}
We study spin dependent transport through a magnetic bilayer
graphene nanojunction configured as two dimensional
normal/ferromagnetic/normal structure where the gate-voltage is
applied on the layers of ferromagnetic graphene. Based on the
fourband Hamiltonian, conductance is calculated by using Landauer
Butikker formula at zero temperature. For parallel configuration
of the ferromagnetic layers of bilayer graphene, the energy band
structure is metallic and spin polarization reaches to its
maximum value close to the resonant states, while for
antiparallel configuration, the nanojunction behaves as a
semiconductor and there is no spin filtering. As a result, a huge
magnetoresistance is achievable by altering the configurations of
ferromagnetic graphene especially around the band gap.
\end{abstract}


 \maketitle
\section{Introduction}
Since spin-orbit coupling in graphene \cite{novoselov} is very
weak \cite{spinorbit} and also there is no nuclear
spin\cite{hyperfine}, spin flip length is so long about $1\mu m$
in dirty samples and room temperature\cite{spinlength}. Clean
samples are expected to have longer spin coherency. This is good
opportunity for spintronic applications based on candidatory of
graphene. On the other hand, graphene has not intrinsically
ferromagnetic (FM) properties, however, it is possible to induce
ferromagnetism externally by doping and defects\cite{defects},
Coulomb interactions \cite{coulomb} or by applying an external
electric field in the transverse direction in
nanoribbons\cite{electric}. Recently, Haugen\cite{Haugen} proposed
the FM correlations due to strong proximity of magnetic states
close to graphene. The overlap between the wave functions of the
localized magnetic states in the magnetic insulator and the
itinerant electrons in graphene induces an exchange field on
itinerant electrons in graphene giving rise spin splitting of the
transport. The exchange splitting which is induced by FM
insulator Euo in graphene was estimated to be of order of $5
meV$. This splitting which is effectively similar to a Zeeman
interaction has so large magnitude that can have important
effects. Such spin splitting can be directly evaluated from the
transmission resonances or magnetoresistance of FM graphene
junction. The ferromagnetism leads to a spin splitting
effectively similar to a Zeeman interaction but of much larger
magnitude. The induced exchange field is tunable by an in-plane
external electric field\cite{ex-tune}. The possibility of
controlling spin conductance in FM monolayer graphene insulator
has also been studied by Yokoyama \cite{Yokoyama}. It was found
that the spin conductance has an oscillatory behavior in terms of
chemical potential and the gate voltage.
\bfig
\includegraphics[width=7 cm]{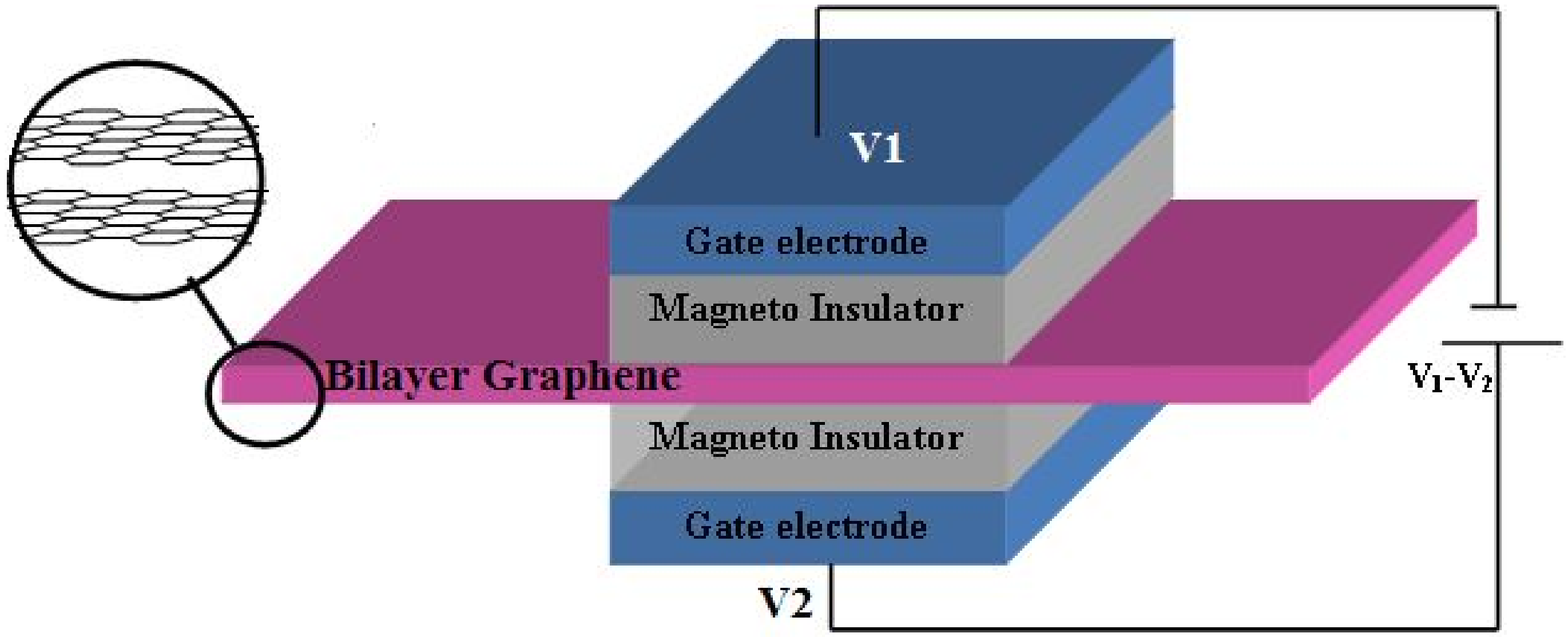}
\caption{Schematic view of normal/ferromagnetic/normal bilayer
graphene. Two gate electrodes can be coated on top of the magnetic
insulator strips which are located on the upper and lower
layers.}\label{model} \efig

Bilayer graphene on the other hand has shown to have interesting
properties for application in nanoelectronic devices such as
transistors based on graphene substrate. The new type of integer
quantum Hall effects \cite{QHE}and also the electronic band gap
controllable by vertically applied electric field are of its
unusual properties in compared to monolayer
graphene\cite{4band,gap-BG,gap-BG1,gap-BG2}. Moreover, the
parabolic band structure close to the Dirac points transforms to
a Mexican-hat like dispersion when an electric field is applied
on graphene. Optical measurements and theoretical predictions
propose a $200$ meV gap in bilayer graphene. This controllable
gap makes bilayer graphene as an appropriate candidate for
spintronic devices. An effective two-band Hamiltonian can
describe the low energy excitations of a graphene bilayer in the
regime of low barrier heights\cite{effective2band}. However, the
four-band Hamiltonian is known to give a better agreement with
both experimental data and theoretical tight-binding
calculations\cite{barbier,4band}. Very recently, spin splitting
of conductance in bilayer graphene has been investigated by using
an effective two-band Hamiltonian emerging from low energy
approximation\cite{Yu}. This approximation is valid when energy
of electrons hitting on the potential barrier is about the
barrier height. Application of  a potential difference between
upper and lower layers intensifies the failure of this
approximation. On the other hand, magnetoresistance in bilayer
graphene has been studied in Ref.\cite{semenov} by using
$8\times8$ Hamiltonian when the induced exchange fields are laid
in plane of each layer with a rotation in their orientations
aginst each other. By using Kobo-Greenwood formula, they have
investigated the dependence of conductivity and
magnetoresistance on temperature and induced exchange field.  

Motivated by these studies, based on the four-band Hamiltonian
and close to the Dirac points, we study spin current through a
magnetic barrier creating by use of the proximity of a
ferromagnetic insulator on bilayer graphene. Conductance is
calculated by use of Landauer-Buttiker formula at zero
temperature. The parameters of the barrier, energy
and angle of incident electrons can affect transport through a
magnetic barrier classifying in the propagating or evanescent
modes. The dependnce of resonant peaks in transmission on system
parameters is proposed to follow a resonance condtion. We have
found in some resonant energies and barrier
parameters and also around the gapped region that a remarkable
spin polarization and also magnetoresistance can be achieved.

{\it Model}: We consider a normal/ferromagnetic/normal bilayer
graphene nanojunction. The model we have used for a bilayer
graphene sandwiched between two ferromagnetic insulator is shown
schematically in Fig.~\ref{model}. The exchange fields induced by
the ferromagnetic insulators is supposed to be perpendicular to
the graphene plane. Therefore, Hamiltonian of the spin up detaches
from the spin down. Two gate electrodes can be attached to the
ferromagnetic graphene from the upper and lower layers which
control the barrier height in each of layers. This set-up is
different from the systems studied by
Refs.\cite{semenov,nguyen}. The exchange field splits this
potential depending on the spin parallel $(+)$ or antiparallel
($-$) to the exchange field. So in the ferromagnetic part of
bilayer graphene, we have $V^{\pm}=V_0 \mp \Delta$ where $\Delta$
and $V_0$ are the exchange field and the potential barrier made
by the gate voltage, respectively. So the two spins are scattered
from the barriers with different heights. It means that energy
shift of the top of the valance band in the barrier is different
for parallel and antiparallel spins to the exchange field. This
spin splitting causes to shift conductance as a function of
energy for each spin resulting in magnetoresistance. To
investigate spin polarization and also magnetoresistance, we have
considered two different configurations so that the exchange
field inducing by the magnetic insulators on each layer are
parallel or antiparallel with respect to each other. The
configuration prepared for observation of magnetoresistance differs from the
configuration considered by Ref.\cite{semenov,nguyen}. The
parallel configuration has a metallic behavior while the
antiparallel configuration induces a potential difference between
upper and lower layers concluding that the system has a
semiconductor behavior with the band gap of $2\Delta$.

This paper is organized as follows: we briefly explain the
formalism which is used for calculating of transmission based on
the four-band Hamiltonian. Before presenting our results, it is so
important to have a short review in section III on transport
through a barrier deposited on bilayer graphene and its
dependence on the system parameters such as energy of incident
quasi-particles and their angle hitting into the barrier and also
barrier parameters. The method presented in section II is a
detailed analysis accompanied with some small corrections on the
method used by Ref.\cite{barbier}. We will present spin
polarization in the parallel configuration in section IV.
Magnetoresistance and its dependence on energy of incident
particles and also induced magnetic field will be investigated in
section V. Finally, the last section concludes our results.
\section{Formalism}
\bfig
\includegraphics[width=6 cm]{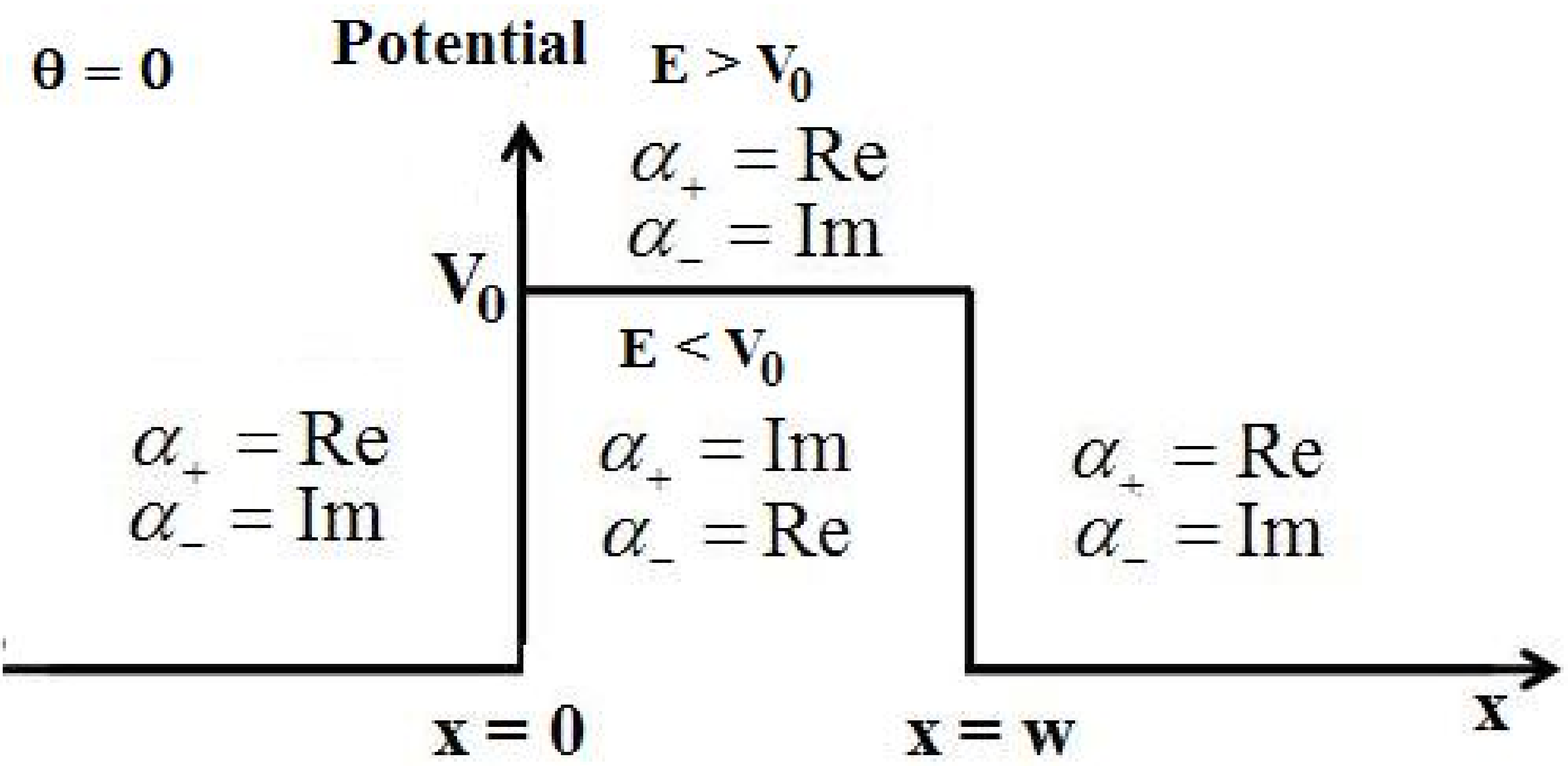}
\caption{Schematic view of the potential barrier with height $V_0$
and width $w$ and the wave number at normal incidence directed in
the x-axis for three regions. }\label{schematic} \efig

In the unit cell of bilayer graphene, we suppose that two
independent sublattices A and B related to each monolayer graphene
are connected to each other in the Bernal stacking. Close to the
Dirac points and in the nearest neighbor tight binding
approximation, the four band Hamiltonian and also its
eigenfunction is written as the following:

\begin{equation}
 H=\begin{pmatrix}V_{1} & \pi & t_{\perp} & 0\\
 \pi ^{\dagger} & V_{1} & 0 & 0\\
 t_{\perp} & 0 & V_{2} & \pi^{\dagger}\\
 0 & 0 & \pi & V_{2}\end{pmatrix}   \,\,\,\, , \,\,\,\,
 \Psi =\begin{pmatrix}\psi _{A}\\
 \psi _{B}\\
 \psi _{B^{\prime}}\\
 \psi _{A^ {\prime}} \end{pmatrix}
\end{equation}

where $$ \pi =\left(p_{x}+ip_{y}\right)v_{f}=-i\hbar v_{f}\left(\partial_{x}-k_{y}\right)$$\\
and in the above formula $ k_{y}=k\sin \theta$ where $ \theta $
and $k$ are the incident angle and wave number of quasi-particles
hitting on a barrier which is created by applying a bias to a
metallic strip deposited on bilayer graphene. Moreover, $V_1$ and
$V_2$ are the gate potentials applied on the upper and lower
layers of bilayer graphene. Such a gate potential can be
manipulated by applying a perpendicular electric field on graphene
sheet. Here, the barrier is approximated by a square potential of
barrier with sharp variation. By solving the eigenvalue equation
of $ H\Psi= E\Psi$, the four band spectrum can be concluded as the
following:

\beq(\varepsilon^{\prime})^2=k^{2}+\delta^{2}+\frac{(t^{\prime})^{2}}{2}
\pm
t^{\prime}\sqrt{4k^{2}\delta^{2}/(t^{\prime})^{2}+(\frac{t^{\prime}}{2})^{2}
+k^{2}}\eeq

where the above parameters are defined as the following:

\begin{eqnarray} \varepsilon^{\prime}=(E-V_0)/\hbar
v_{f}=\varepsilon-\upsilon_0 , \,\,\,\,\ V_0=(V_1+V_2)/2 \nonumber
\\  \delta=(V_{1}-V_{2})/2\hbar v_{f}, \,\,\,\,\,\
t^{\prime }=t/\hbar v_{f}. \label{parameters}
 \end{eqnarray}
In the case of $\delta=0$ and $k<<t^\prime$,  in low energy limit, energy spectrum behaves as $E-V_0=\pm \hbar^2k^2/2m$ where $m=t/(2v_f^2)$ is an effective mass. This approximation which results in a effective two-band Hamiltonian is valid when energy of incident electrons is close to the barrier height. On the other word, in the case of zero potential difference between two layers, the absolute value of $(E-V_0)$ should be much smaller than the interlayer coupling strength (0.4eV). For $\delta \neq 0$, this approximation may fail for large potential differences. So one should care to choose valid energy and potential ranges. However, in spite of Ref.\cite{Yu}, in this paper, we use four-band Hamiltonian in which the only approximation is the Dirac cone.

If we assume plane wave solution for the Schroedinger equation,
the wave function in each region with a constant potential is
written as the following matrix product.

\begin{equation}
\Psi= GM \begin{pmatrix} a\\b\\c\\d
\end{pmatrix}
\label{wavefunction}\end{equation} where matrix elements of the
matrices $G$ and $M$ are defined as the following:
$$ G=\begin{pmatrix}
1&1&1&1\\
f_{+}^+& f_{-}^+ & f_{+}^- & f_{-}^- \\
h^{+}& h^{+} & h^{-}&h^{-}\\
g_{+}^+h^{+} & g_{-}^+h^{+} & g_{+}^-h^{-} & g_{-}^- h^{-}
\end{pmatrix}$$
\beq M(x)=\begin{pmatrix}
e^{i\alpha_{+}x} & 0 & 0 & 0\\
0 & e^{-i\alpha_{+}x} & 0& 0 \\
0 & 0 & e^{i\alpha_{-}x} & 0\\
0& 0 & 0 & e^{-i\alpha_{-}x}
\end{pmatrix}
\eeq

\begin{equation}
\begin{array}{c}
f_{\pm}^{+}=\dfrac{\pm\alpha_{+}-ik_{y}}{\varepsilon
^{\prime}-\delta} \,\,\,\,\, ,\,\,\,\,\,
f_{\pm}^-=\dfrac{\pm\alpha_{-}-ik_{y}}{\varepsilon ^{\prime}-\delta}  \\

g_{\pm}^+=\dfrac{\pm\alpha_{+}+ik_{y}}{\varepsilon^{\prime}+\delta}
\,\,\,\,\, ,\,\,\,\,\,
g_{\pm}^-=\dfrac{\pm\alpha_{-}+ik_{y}}{\varepsilon^{\prime}+\delta}\\

h^{\pm}=\dfrac{(\varepsilon^{\prime}-\delta)^{2}-\alpha_{\pm}^{2}-k_{y}^{2}
} {t^{\prime}(\varepsilon^{\prime}-\delta) }
\end{array}
\end{equation}\\
Here $\alpha$ is the wave number in the x direction and is
defined as:

\beq \alpha_{\pm}^{2}
=[\delta^{2}+(\varepsilon^{\prime})^{2}-k_{y}^2\pm
\sqrt{4(\varepsilon^{\prime})^{2}\delta^{2}+
(t^{\prime})^{2}((\varepsilon^{\prime})^{2} -\delta^{2})}]
\label{wavenumber}
 \eeq
In the special case of {\it normal incident angle} and zero gate
potential where $k_y=\delta=0$, the wave number $\alpha_{+}$ is
real in the energy range of $0<\varepsilon^{\prime}<t^{\prime}$
and $\alpha_{-}$ is real if the energy of incident particles is
in the range of
$\varepsilon^{\prime}<0,\varepsilon^{\prime}>-t^{\prime}$. In
this paper, studied system contains a magnetic or electrostatic
barrier of potential as shown in Fig.\ref{schematic}. The barrier
width is $w$. The electrostatic potential which plays the role of
the gate voltage is set to be $V_0$ in the barrier part and zero
in the first and last regions. We suppose that the energy range
of incident particles is limited to the range of
$0<\varepsilon^{\prime}<t^{\prime}$. Consequently in the barrier
part we have $-V_0<\varepsilon^{\prime}_2<t^{\prime}-V_0$. The
wave numbers behind and in front of the barrier $\alpha_+^{(1)}$
and $\alpha_+^{(3)}$ are real while $\alpha_-^{(1)}$ and
$\alpha_-^{(3)}$ are imaginary. In the barrier part, for
$\varepsilon^{\prime}_2<0$, the wave numbers $\alpha_+^{(2)}$ and
$\alpha_-^{(2)}$ are imaginary and real, respectively, while for
$\varepsilon^{\prime}_2>0$ they behave vice versa. A schematic
view of the barrier at normal incidence and wave numbers in each
part are shown in Fig.\ref{schematic}.
\bfig
\includegraphics[width=9 cm]{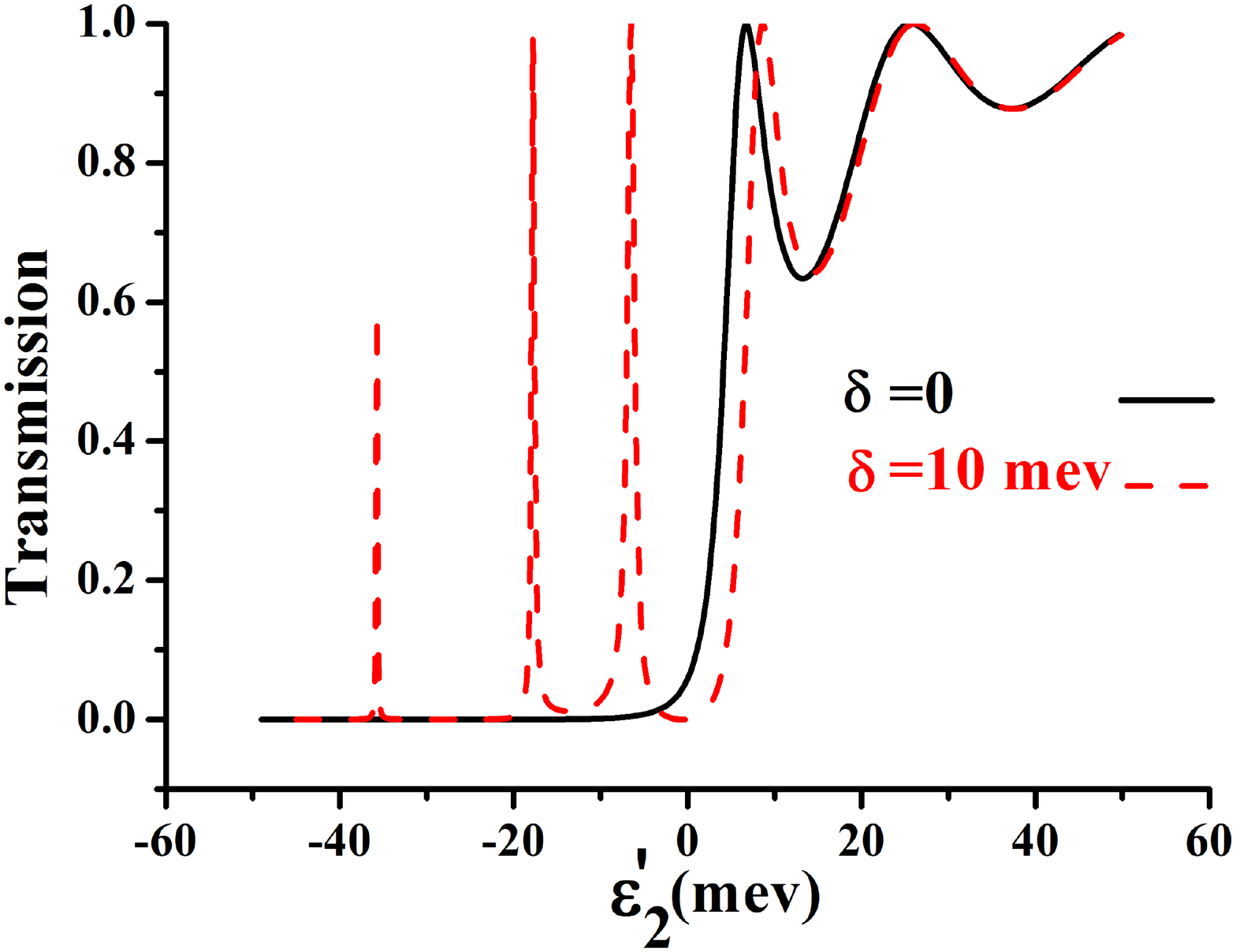}
\caption{Transmission at normal incidence $\theta=0$ as a
function of energy difference between energy of incident
particles and the barrier height $(E-V_0)/\hbar v_F$ for potential
difference of upper layer against lower layer $\delta=0$ and $10
meV$.} \label{normalincident}\efig

By applying continuity of the wave functions on the boundaries of
the barrier, one can connect coefficient matrix of the wave
function for the last region $A_3$ to the coefficient matrix for
the first region $A_1$. \beq
 \begin{array} {c}
A_{1}=NA_{3}  \\
N=M_{1}^{-1}(0)G_{1}^{-1}G_{2}M_{2}(0)M_{2}^{-1}(w)G_{2}^{-1}G_{3}M_{3}(w)
\label{transfermatrix}\end{array} \eeq where N is called as
transfer matrix. Since $\alpha_{-}^{(1)}$ and $\alpha_{-}^{(3)}$
are imaginary in the interested energy range, that part of wave
function which are associated by such wave numbers are
exponentially a growing or decaying function. So we have to set
the coefficient of plane wave $e^{i \alpha_{-}^{(1)}x}$ ($c$ in
Eq.~\ref{wavefunction}) to be zero for the first region, because
this part of wave function grows exponentially when
$x\rightarrow-\infty$. Therefore, coefficients matrix in the
first region is supposed to be as $A_1=[1,r,0,e_g]^T$, where the
superscript $T$ refers to the transpose of a matrix and $e_g$ is
the coefficient of growing evanescent state and $r$ is the
coefficient of the reflected part of the wave function.  In the
last region, we have to set the coefficient ($d$ in
Eq.~\ref{wavefunction}) of $e^{-i \alpha_{-}^{(3)}x}$ to be zero
because this part of the wave function increases exponentially
when $x\rightarrow\infty$. Therefore, the coefficients matrix in
the last regions is supposed to be as $A_3=[t,0,e_d,0]^T$ where
$t$ is the coefficient of the transmitted part of wave function
and $e_d$ is the coefficient of decaying evanescent state. In
this region, there is no reflected wave. However, in equation 8 of Ref.\cite{barbier},
matrices $A_1$ and $A_3$ have been considered to be completely displaced which leas to different results.

By rearrangement of the transfer matrix elements of Eq.~\ref{transfermatrix}, the
coefficient of transmitted part of wave function is derived in
terms of transfer matrix elements as the following;

 \beq
t=[N_{11}-N_{13}N_{31}/N_{33}]^{-1}.\eeq Since the first and last
regions possess similar wave numbers, transmission probability is
given as $T=|t|^2$. Before presenting our results, in the next
section, we will shortly review transport properties through a
potential barrier by using the mentioned formalism.
\bfig
\includegraphics[width=9 cm]{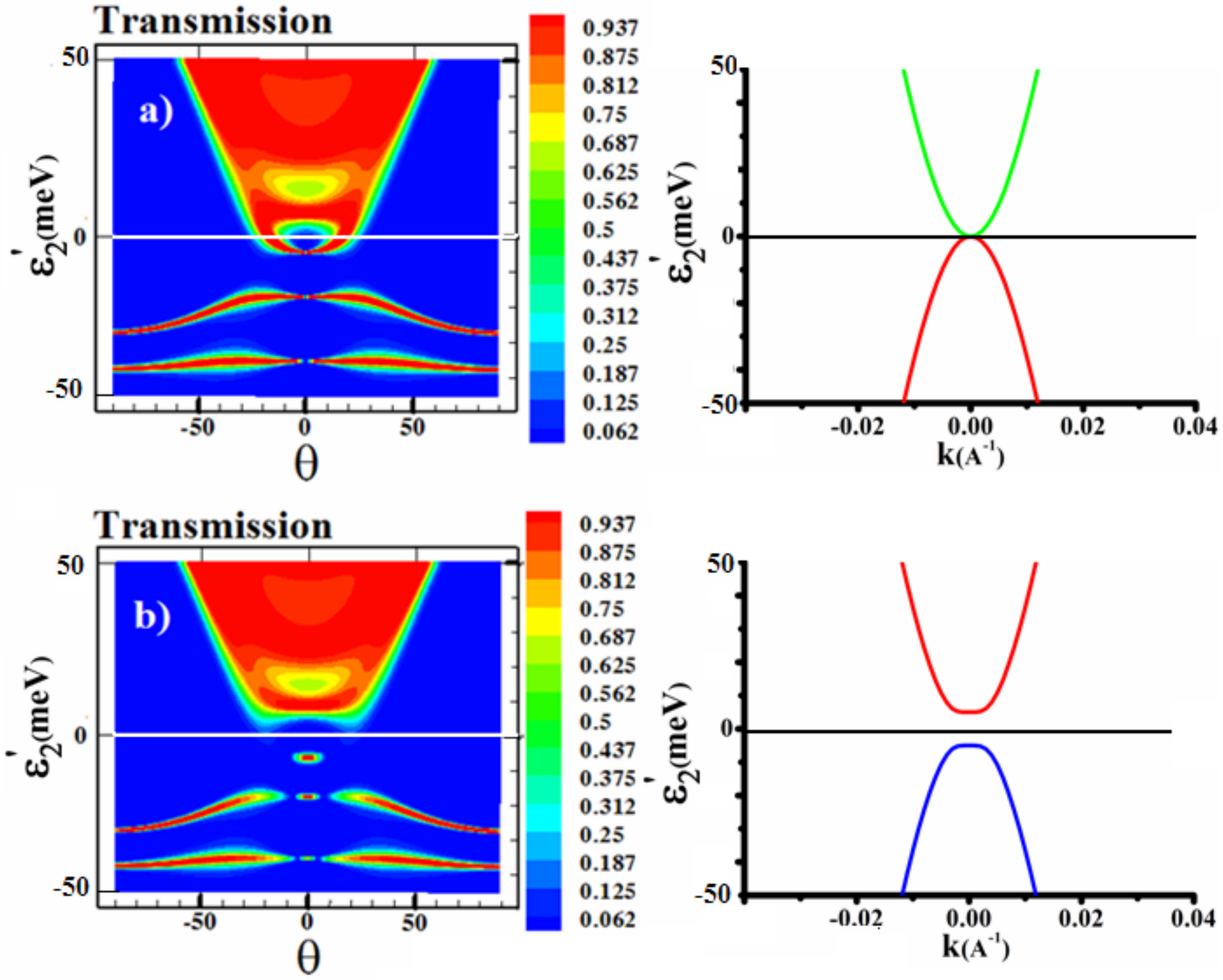}
\caption{Contour plot of transmission in plane of the incident
angle and energy difference of
$\varepsilon^{\prime}_2=(E-V_0)/\hbar v_F$ accompanied with the
band structure for potential difference of the upper layer against
the lower layer to be as a) $\delta=0$ and b) $10 meV$.}
\label{contourplotT}\efig

\section{Transmission through a Barrier on Bilayer Graphene}

The Klein tunneling in monolayer graphene results in a complete
transmission through a barrier potential in normal incident.
However, in contrast to monolayer graphene, as a result of chiral
symmetry in bilayer graphene, transmission is zero for
quasiparticles with energies lower than the barrier height. In
the special case of $\delta=k_y=0$, transmission through a
potential barrier can be analytically calculated in normal
incident and for two ranges of energy $\varepsilon^{\prime}<0$
and $\varepsilon^{\prime}>0$.
\begin{equation}
\begin{array}{c}
t=e^{i\alpha^{(1)}w}[\cos(\alpha^{(2)}w)-iQ\sin(\alpha^{(2)}w)]^{-1} \\
\end{array}
\end{equation}
where $$Q=\frac {1}{2}(\frac
{\varepsilon^{\prime}_{1}\alpha^{(2)}}{\varepsilon_{2}^{\prime}\alpha^{(1)}}
+\frac{\varepsilon _{2}^{\prime} \alpha ^{(1)}}{\varepsilon
_{1}^{\prime}\alpha ^{(2)}})$$ where in the above formula,
parameters are defined as $ \varepsilon_{1}^{\prime}
=\varepsilon/\hbar v_{f} , \varepsilon_{2}^{\prime}
=(\varepsilon-V_0)/\hbar v_{f}$. So the real part of the wave
numbers inside and outside of the barrier part are defined as
$\alpha^{(1)}=[(\varepsilon_{1}^{\prime})^{2
}+\varepsilon_{1}^{\prime} t^{\prime} ]^{1/2}$ and $\alpha^{(2)}=
 [(\varepsilon_{2}^{\prime})^{2}+\varepsilon^{\prime}_{2}t^{\prime}]^{1/2}$,
 respectively.  The energy of incident particle is supposed to be
 always $\varepsilon^{\prime}_1>0$. So $\alpha^{(1)}$
 is always real.

 For the energy range of $E<V_0$, the
 wave number inside the barrier part $\alpha^{(2)}$ and consequently $Q$ are imaginary
 so that $\alpha^{(2)}=i\kappa$ and $Q=iq$ where $\kappa$ and $q$
 are real. As a result, transmission tends to zero as a function
 of the system parameters such as $w$ and $\varepsilon^{\prime}_2$.

 \begin{equation}
T(\theta=0,\varepsilon^{\prime}_2 < 0)
=tt^{\ast}=[\cosh^{2}(\kappa w)+q^{2}\sinh^{2}(\kappa w)]^{-1}
\end{equation}
This behavior is the trace of chiral symmetry in bilayer
graphene. However, if the incident angle is nonzero, some resonant
peaks appear in the transmission curve (see
Fig.~\ref{contourplotT}). For the energy range of $E>V_0$, all
parameters such as $\alpha^{(2)}$ and $Q$ are real. Thus
transmission has an oscillatory behavior as a function of
$\varepsilon^{\prime}_2$ as the following form,

\beq T(\theta=0,\varepsilon^{\prime}_2 >
0)=[\cos^{2}(\alpha_{+}^{(2)} w)+Q^{2}\sin^{2}(\alpha_{+}^{(2)}
w)]^{-1} \eeq
\bfig
\includegraphics[width=6 cm]{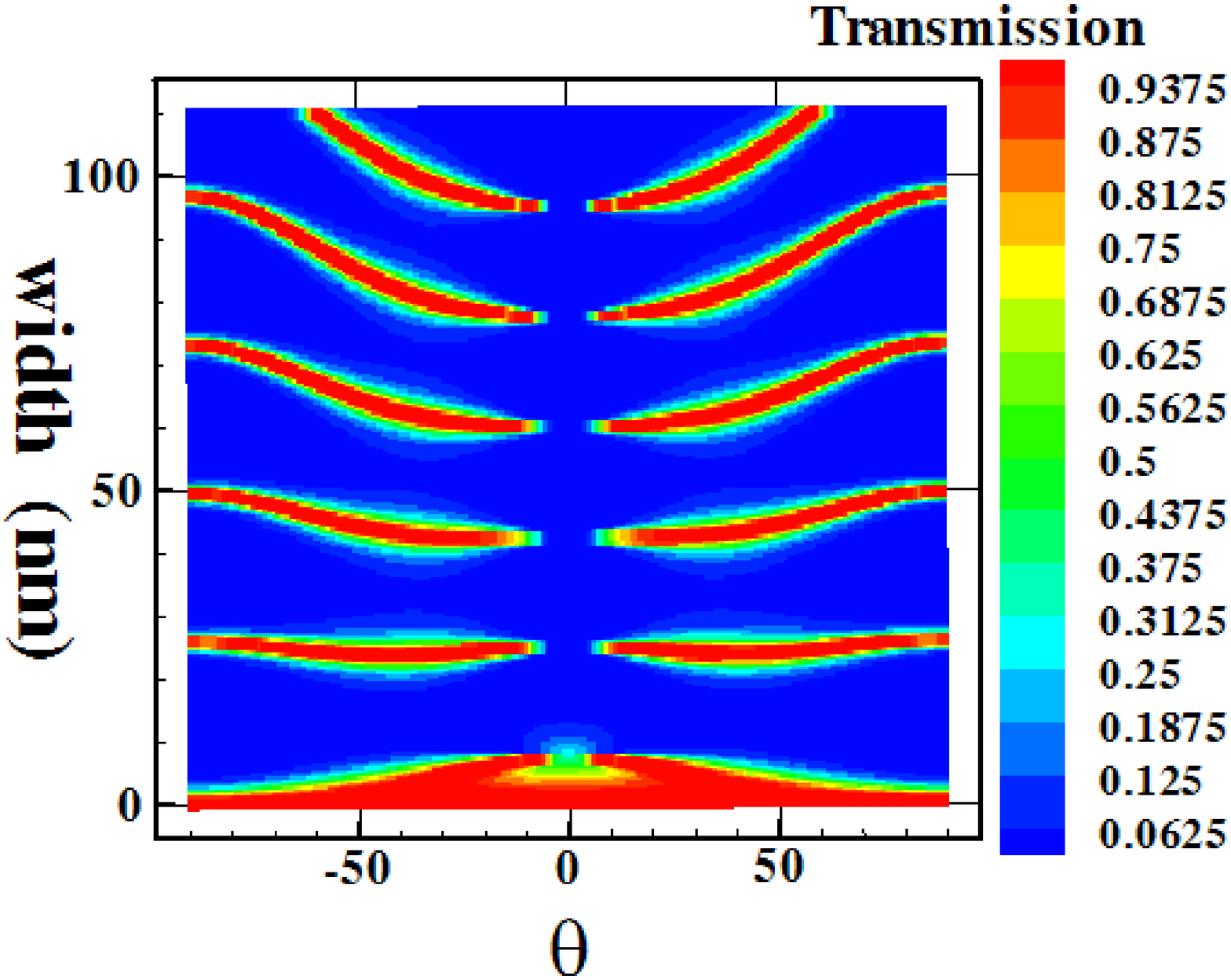}
\caption{Contour plot of transmission in plane of the incident
angle and the barrier width for the barrier height $50 meV$, the
energy of incident angle $17 meV$ and for the case of $\delta=0$.}
\label{contourplotwidth}\efig

In the high energies limit $E\gg V_0$, we have $Q \longrightarrow
1$ and so transmission is complete ($T \longrightarrow 1$). By
applying a vertically electric field in the barrier part, a band
gap is opened in the band structure of bilayer graphene which is
proportional to the potential difference between potentials of
each layers. In this case, chiral symmetry is failed and therefore
transmission in normal incidence is nonzero for energies lower
than the barrier height ($E<V_0$). Transmission at normal
incidence is represented in Fig.~\ref{normalincident} as a
function of $\varepsilon^{\prime}_2$ for $\delta=0$ and $10 meV$.
Application of a vertically electric field causes to emerge some
resonant tunneling states for energies of $E<V_0$. In this energy
range, resonant states originates from interference of the
incident and scattered waves. For all cases such as nonzero
incident angles and $\delta\neq0$, the resonant peaks are
interpreted by the resonance condition relation proposed as

\beq \alpha_{b}w=n\pi \label{resonance}\eeq
 where $\alpha_b$ is the x-component of the
wave number inside the barrier which can be calculated by
Eq.~\ref{wavenumber}.

To have more complete view, we prepare a contour plot of
transmission in plane of the incident angle and
$\varepsilon^{\prime}_2$ which is shown in
Fig.~\ref{contourplotT} for a fixed width of the barrier. For the
normal incidence ($\theta=0$), transmission behavior is
compatible with the results shown in Fig.~\ref{normalincident}.
\bfig
\includegraphics[width=9 cm]{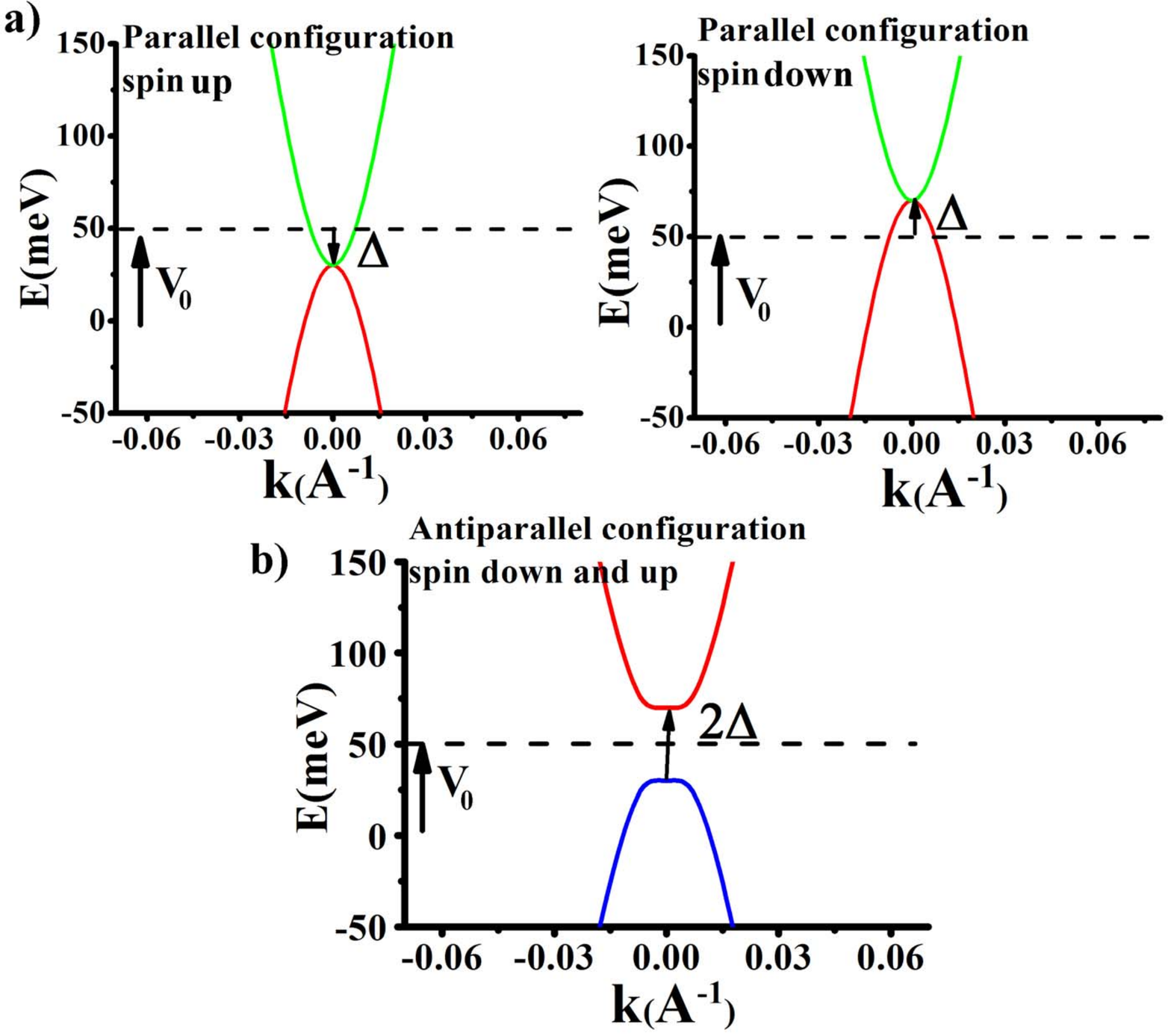}
\caption{Energy band structure for parallel and antiparallel
configurations. In the parallel configuration, the direction of
the exchange fields inducing in each layers are parallel with
respect to each other. a) for spin parallel and antiparallel to
the exchange field direction in the parallel configuration. b)
for both spins up and down in the antiparallel configuration.}
\label{bands}\efig

For energies higher than the barrier height
$\varepsilon^{\prime}_2>0$, transmitting channels are opened over
all ranges of energies. However, transmitting window for the
incident angles is limited with the condition that
$\alpha_+^{(2)}$ (in Eq.~\ref{wavenumber}) is real. In the case
of $\delta=0$, the range of incident angle in which transmission
is high can be extracted as $-\sin^{-1}
\frac{\sqrt{(\varepsilon^{\prime}_2)^2+\varepsilon^{\prime}_2
t^{\prime}}}{k}\leq \theta \leq \sin^{-1}
\frac{\sqrt{(\varepsilon^{\prime}_2)^2+\varepsilon^{\prime}_2
t^{\prime}}}{k}$. Therefore, by increasing
$\varepsilon^{\prime}_2$, the range of angles with high
transmission becomes more extended. In the energy range of
$\varepsilon^{\prime}_2<0$, independent of the value of $\delta$,
resonant peaks emerge for nonzero incident angles ($\theta\neq
0$) which obey the resonance condition $\alpha_{b}w=n\pi$. So
additional to some resonant energy states, we have some resonant
widths in which transmission is high. Fig.~\ref{contourplotwidth}
shows transmission in plane of the incident angle and the barrier
width for $\varepsilon^{\prime}<0$ and $\delta=0$. It is shown
that based on the resonance condition
(Eqs.~\ref{resonance},\ref{wavenumber}), in large incident
angles, $\alpha_{b}$ reduces and so in a fixed resonance order
($n$), the resonance condition is satisfied for wide barriers.
Therefore, the resonance strips with complete transmission shown
in Fig.~\ref{contourplotwidth}, depend strongly on the incident
angle in the range of wide barriers.

By applying a vertically electric field in the barrier part, a
band gap is opened around $\varepsilon^{\prime}_2=0$. This band
gap also has a trace in transmission as a transport gap shown in
Fig.~\ref{contourplotT}b.

\bfig
\includegraphics[width=9 cm]{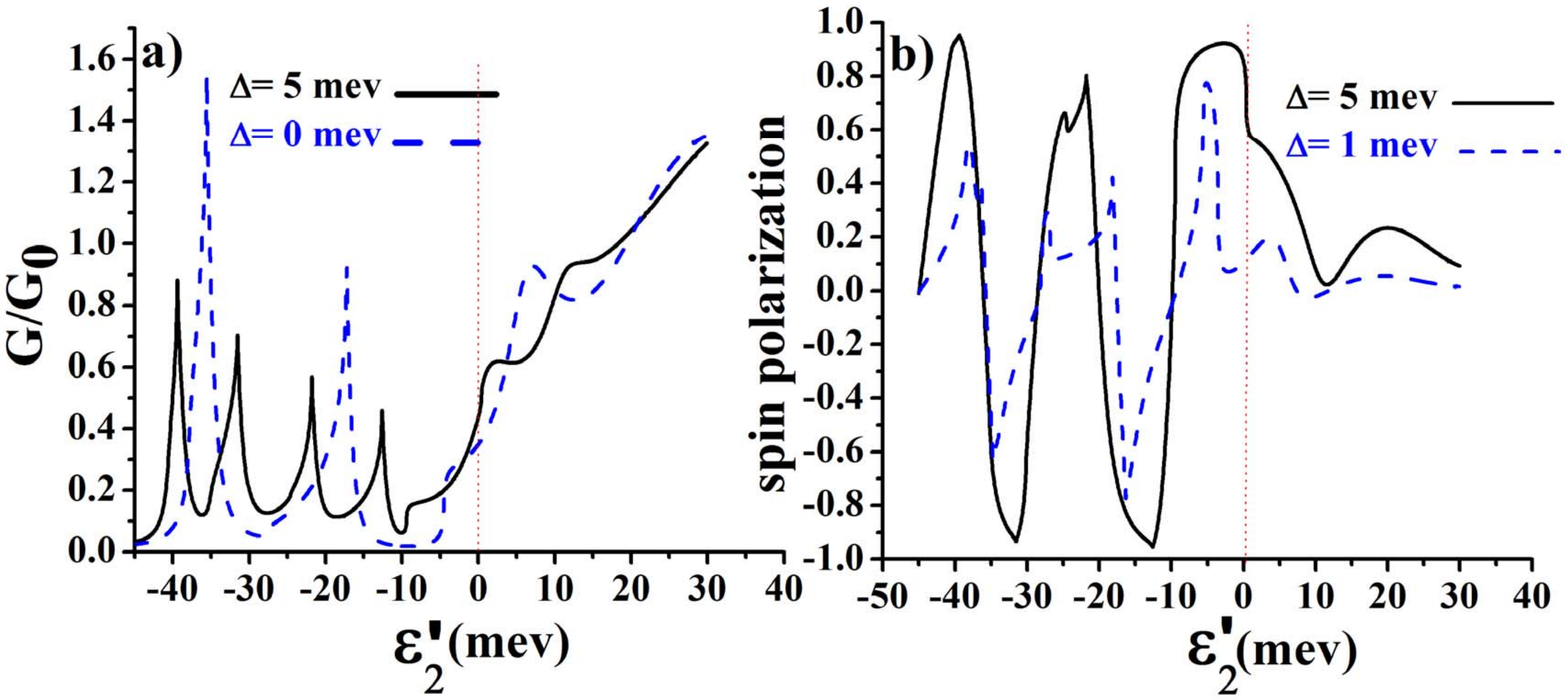}
\caption{a) Conductance and b) spin polarization as a function of
$\varepsilon^{\prime}_2$ for different induced exchange fields
$\Delta$ in the parallel configuration. Here, the barrier height
and width are considered to be as $50$ meV and $40$ nm,
respectively.} \label{spinpol}\efig
\section{Results}
By application of an averaged gate voltage $V_0$, band structure
in the barrier part is shifted by $V_0$ value.  Fig.~\ref{bands}
shows band structure of parallel and antiparallel configuration
magnetic insulators when a gate voltage is applied on the barrier
part. In case the exchange fields inducing in each layers of
bilayer graphene are parallel, particles with spin parallel (spin
up) and antiparallel (spin down) to the exchange fields are
scattered from barriers with different heights. In the parallel
configuration, spin splitting of the barrier potential in the
ferromagnetic graphene is written as $V^--V^+=2\Delta$. Such spin
splitting is also seen in the band structure that is shown in
Fig.~\ref{bands}a. It is seen that the top of valance band are
shifted to lower/higher energies for spins up/down. However, in
the antiparallel configuration, the band structure shown in
Fig.~\ref{bands}b is the same for both up and down spins. A band
gap which is proportional to $2\Delta$ appears in the band
structure of the antiparallel configuration.

\subsection{Spin Polarization}
Here, there is a correspondence between the band structure and
transmission. According to the band structure, we expect to
emerge spin polarization just for parallel configuration because
energy bands for up and down spins are shifted by $2\Delta$ with
respect to each other. However, since the band structure for
antiparallel configuration is the same for both spins, it is not
expected to have spin polarization for this configuration. The
spin polarization is defined as:
\bfig
\includegraphics[width=9 cm]{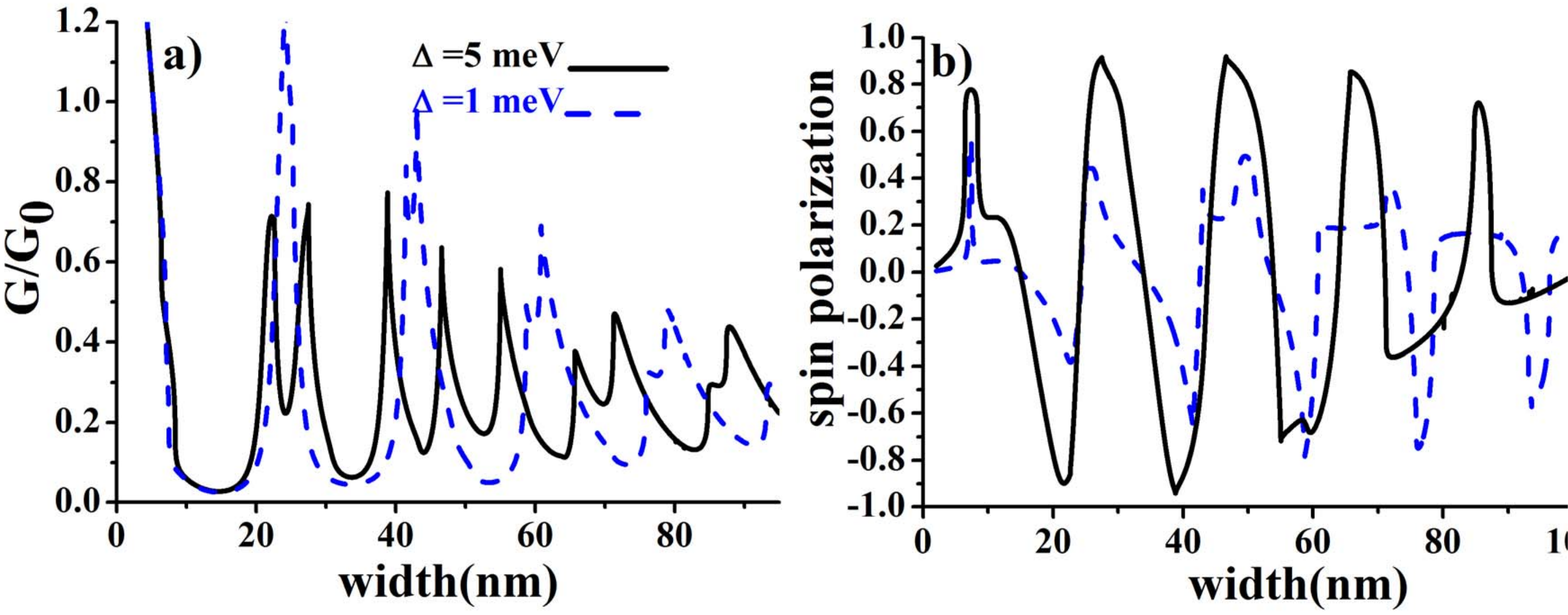}
\caption{a) Conductance and b) spin polarization as a function of
barrier width for different induced exchange fields $\Delta$ in
the parallel configuration. Here, the barrier height and incident
energy are considered to be as $50$ meV and $17$ meV,
respectively.} \label{widthpol}\efig

\begin{equation}
P=\dfrac{G_{up}-G_{down}}{G_{up}+G_{down}} \label{polarization}
\end{equation}
where $G_{up}$ and $G_{down}$ are conductance for up and down
spins. The conductance is calculated by using Landauer formalism
in the linear regime. Therefore, conductance is proportional to
angularly averaged transmission projected along the current
direction.

$$G=\int_{-\pi/2}^{\pi/2}T(E,cos(\theta))cos\theta d\theta $$

It is clear that additional to the transmission curves
(Fig.\ref{contourplotT}), resonance peaks also appears in
conductance. Since up and down spins in the parallel configuration
see barriers with different heights, resonance peaks in
conductance as a function of Fermi energy $E$ are shifted to
higher energies as $\Delta$ for spin down and to lower energies
as $-\Delta$ for spin up. This mismatching of conductance peaks
for two spins causes to a large spin polarization at resonance
states. Fig.~\ref{spinpol} displays conductance and spin
polarization as a function of $\varepsilon^{\prime}_2$ for the
parallel configuration. It is shown that conductance peaks and
consequently spin polarization appears in the energy range of
$\varepsilon^{\prime}_2<0$. It is seen that by inducing an
exchange field, conductance peaks in Fig.~\ref{spinpol}a split
into two peaks which are related to each spin. This spin
splitting is about $2\Delta$. Spin polarization shown in
Fig.~\ref{spinpol}b has an oscillatory behavior with energy of
incident particles for energies lower than the barrier height
$\varepsilon^{\prime}_2<0$. The amplitude of spin polarization
increases with the induced exchange field $\Delta$ and reaches to
its maximum value. However, spin polarization tends to zero for
energies greater than the potential height
$\varepsilon^{\prime}_2>0$ except at $E \sim V_0$.

In the parallel configuration and for $\varepsilon_2^{\prime}<0$,
Fig.~\ref{widthpol}a shows that conductance in the resonance
widths has a peak. These peaks which are also seen in the
transmission curves of Fig.~\ref{contourplotwidth} are explained
by the resonance condition of Eq.~\ref{resonance}. It is shown
that spin splitting of conductance peaks also appears in the
resonance widths which is originated from different barrier
heights for two spins up and down. It should be noted that the
conductance at resonance widths decreases for wide barriers. In
the wide range of widths, the angularly window for transmitting
channels shown in Fig.~\ref{contourplotwidth} decreases with the
widths.
\begin{figure}
\centering
\includegraphics[width=8 cm]{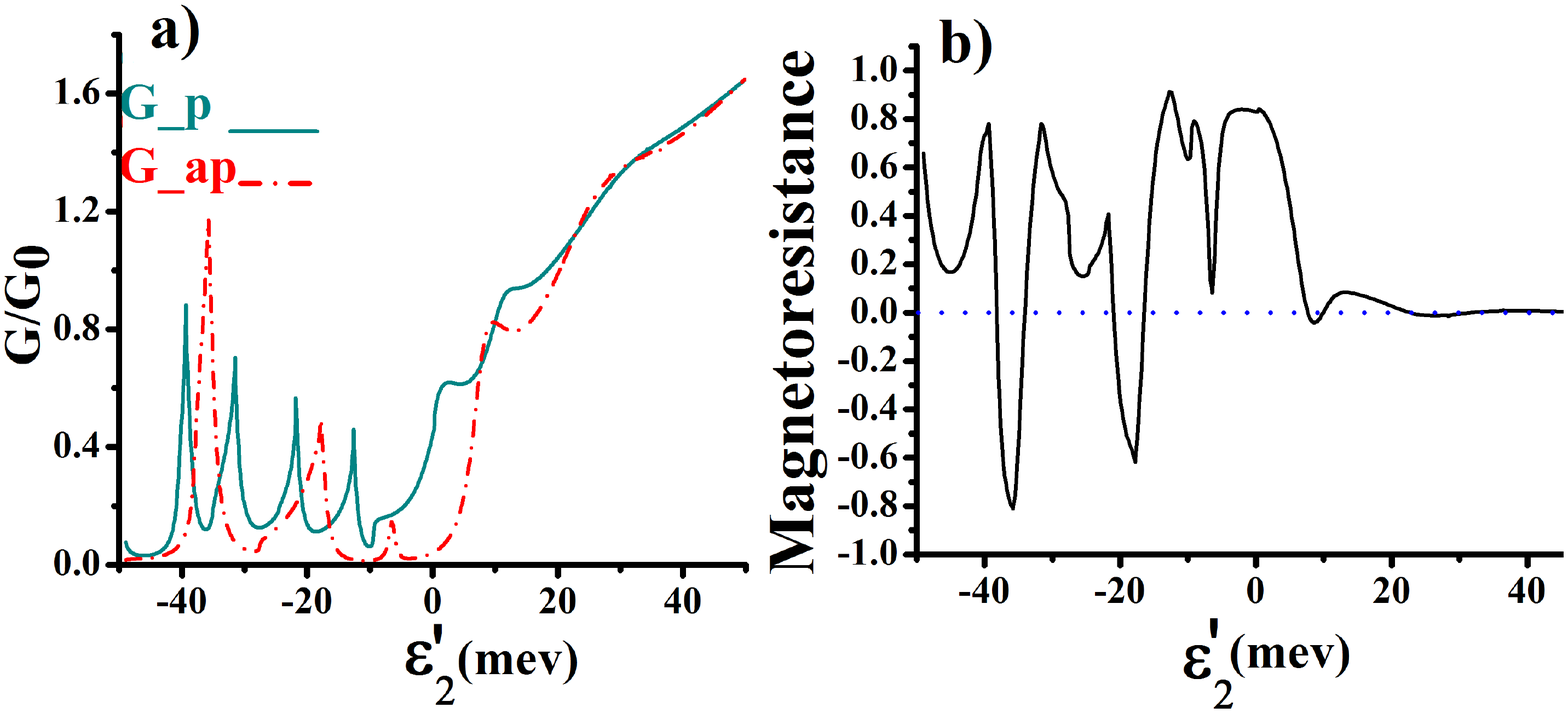}
\includegraphics[width=8 cm]{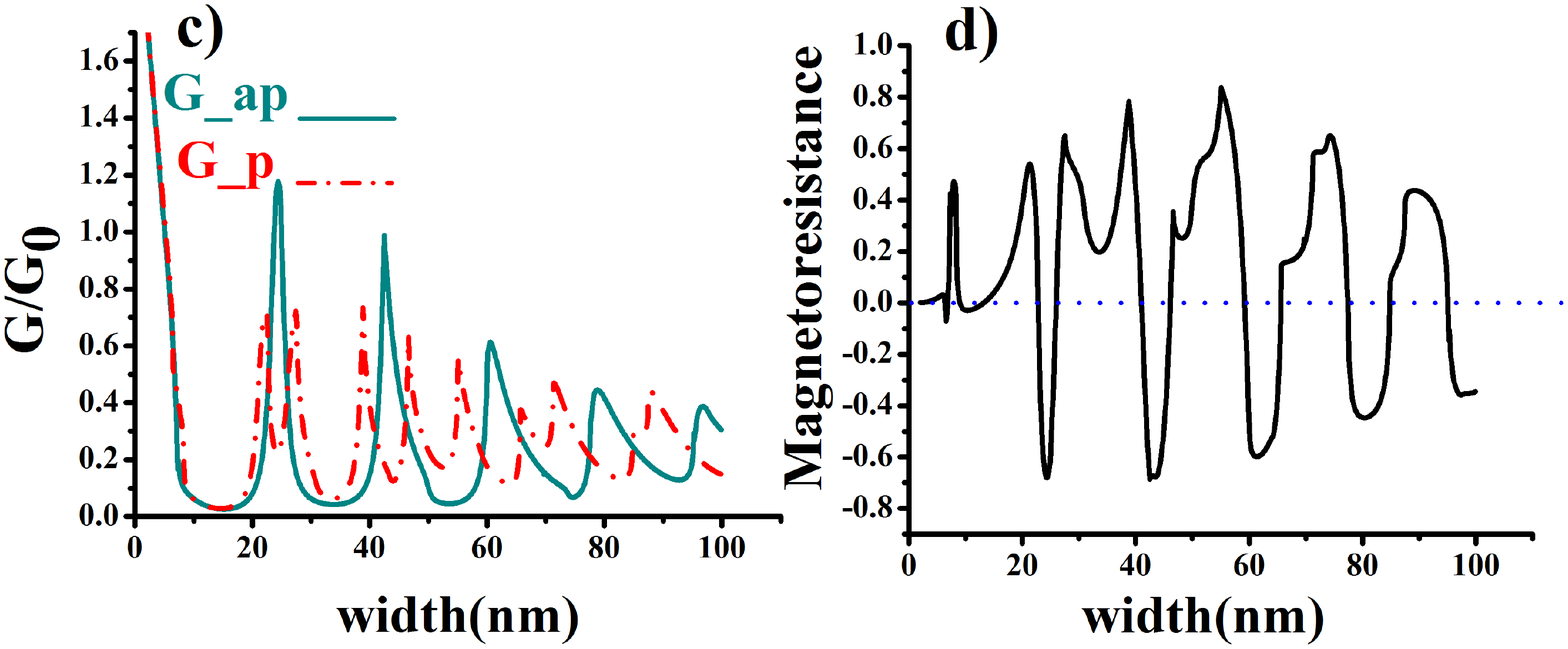}

\caption{Conductance in the parallel and antiparallel
configurations as a function of
a)$\varepsilon^{\prime}_2=(E-V_0)/\hbar v_F$ for a barrier with
the width of 40 nm and, c) the barrier width for a barrier with
the height of 50 meV. Magnetoresistance as a function of b)
$\varepsilon^{\prime}_2$ for a barrier with the width of 40 nm,
d) the barrier width for a barrier with the height of 50 meV. The
induced exchange field is considered to be as $\Delta=5 $meV. }
\label{MR-EW}
\end{figure}

Fig.~\ref{widthpol}b shows spin polarization as a function of the
barrier width. Again, spin polarization has an oscillatory
behavior with the barrier width. The amplitude of spin
polarization strongly increases by an increase of the induced
exchange field. Therefore, to manifest this spin polarization, we
should manufacture the ferromagnetic graphene part with the
special widths in which spin polarization reaches to the value of
unity.

\subsection{Magnetoresistance}
In this section, we will show that by switching between parallel
and antiparallel configurations, one can obtain large
magnetoresistance. Magnetoresistance is defined as the following:

\begin{equation}
MR=\dfrac{G^{p}-G^{ap}}{G^{p}+G^{ap}}
\end{equation}
where $G^{p}=G_{up}^p+G_{down}^p$ and
$G^{ap}=G_{up}^{ap}+G_{down}^{ap}$ are conductance for parallel
and antiparallel configurations.

Fig.~\ref{MR-EW} displays conductance in the parallel and
antiparallel configurations and also magnetoresistance as a
function of $\varepsilon^{\prime}_2$ and the barrier width for a
fixed exchange field $\Delta=5 meV$. As we before expressed, spin
splitting at the resonance states (for
$\varepsilon^{\prime}_2<0$) emerges in conductance peaks in the
case of the parallel configuration. This behavior is clear in
Fig.~\ref{MR-EW}a and \ref{MR-EW}c. However, this splitting will
not occur for the case of antiparallel configuration. Therefore,
large magnetoresistance appears around the conductance resonance
peaks. In the parallel configuration, a band gap appears around
the barrier edge in the interval $V_0-\Delta<E<V_0+\Delta$. This
band gap has a trace in transmission and consequently
conductance. Zero conductance region around the barrier edge
$\varepsilon^{\prime}_2\sim 0$ which is seen in Fig.\ref{MR-EW}a,
is a result of the band gap. Since there is no such a band gap in
the parallel configuration, magnetoresistance as shown in
Fig.\ref{MR-EW}b reaches to its maximum value in the energy band
gap. In the energy range greater than the barrier height
$\varepsilon^{\prime}_2>0$, there is no spin splitting and
therefore, magnetoresistance tends to zero.

As we showed before, conductance has peak at resonant widths.
Similar to the previous case, spin splitting occurs just for the
parallel configuration. So magnetoresistance increases around the
resonance widths. The oscillatory behavior of magnetoresistance
as a function of the barrier width is represented in
Fig.\ref{MR-EW}d.

\begin{figure}
\centering
\includegraphics[width=9 cm]{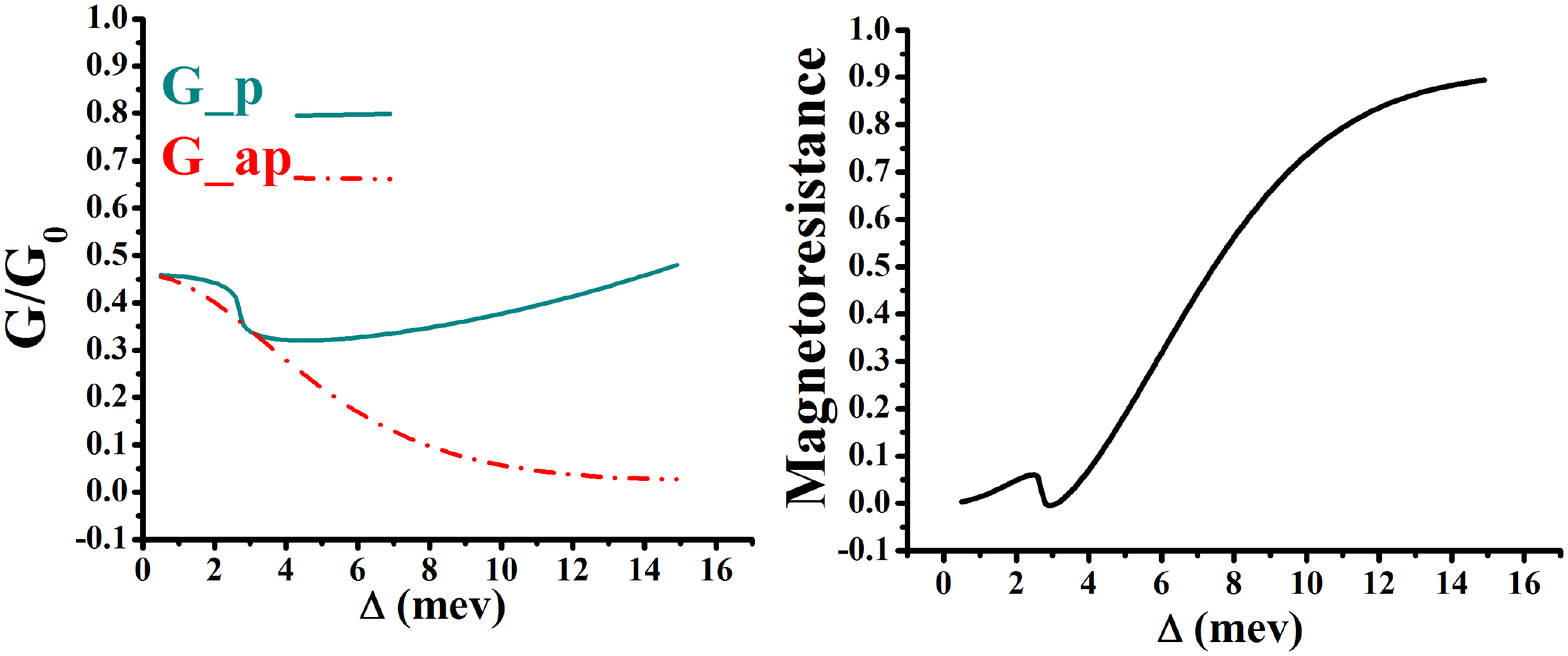}

\caption{a) Conductance in the parallel and antiparallel
configuration and b) magnetoresistance as a function of the
induced exchange field $\Delta$ for a barrier with the height of
50 meV and energy of incident particles as 40 meV. Here the
barrier width is 20 nm.} \label{MR-Delta}
\end{figure}
As we showed, there is a large magnetoresistance around the
barrier edge $E \approx V_0$. In this range of energies, we
investigate the dependence of magnetoresistance to the induced
exchange field. This exchange field of graphene can be controlled
by an in-plane external electric field \cite{ex-tune}.
Fig.~\ref{MR-Delta}b represents that magnetoresistance increases
monotonically by increasing the exchange field $\Delta$. It is
interesting by increasing the exchange field up to $10 meV$,
magnetoresistance reaches to its maximum value.

To explain this behavior, we investigate the dependence of
conductance on the exchange field in the parallel and
antiparallel configurations. In the antiparallel configuration,
the band gap which is limited in the interval of
$V_0-\Delta<E<V_0+\Delta$, enhances by increasing the exchange
field. Therefore, conductance in the antiparallel configuration
goes to zero when the exchange field is increased. Suppression of
the conductance with the exchange field in the antiparallel
configuration is shown in Fig.~\ref{MR-Delta}a. However, in the
parallel configuration, conductance increases by enhancement of
the exchange field. The reason of this enhancement comes back to
have larger angularly transmitting windows for larger
$\varepsilon^{\prime}_2$ (see Fig.~\ref{contourplotT}). In fact,
effective potential for spins up $V^+=V_0-\Delta$ is decreased by
an increase in $\Delta$. So $\varepsilon^{\prime}_2=(E-V_0)/\hbar
v_F$ for a fixed energy is increased and consequently $G_{up}$
and so $G^p$ is increased by $\Delta$. As a conclusion, for the
exchange fields up to 10 meV, suppression of $G^{ap}$ and an
increase of $G^{p}$ results in a large magnetoresistance which is
so useful for designing spin memory devices.

\section{Conclusion}
We have studied spin polarization and magnetoresistance of a
normal/ferromagnetic/normal junction of bilayer graphene by using
transfer matrix method and based on the four-band Hamiltonian.
Transport properties simultaneously is controlled by two gate
electrodes ($V_0$), which are applied on the ferromagnetic
graphene. Two configurations of the exchange field is considered
perpendicular to the graphene sheet. This exchange field is
induced by the proximity of a localized magnetic orbital in a
magnetic insulator coating on top of each layers of bilayer
graphene in the barrier part. In the parallel configuration which
graphene has a metallic behavior, a spin splitting $2\Delta$
occurs for the conductance at the resonant states just for
energies lower than the barrier height $E<V_0$. However, there is
no spin splitting in the antiparallel configuration. A band gap
of $2\Delta$ is opened in the antiparallel configuration which
makes it a semiconductor. As a result of spin splitting in the
parallel configuration, an oscillating spin polarization emerges
for energies lower than the barrier height. Furthermore, an
oscillatory of magnetoresistance with large amplitude is
achievable for $E<V_0$ when we are able to switch between two
configurations. There is also a large magnetoresistance in the energy range around the
barrier edge originating from the band gap which is openned by a vertically electric field. In this range of energy, magnetoresistance reaches to its maximum value
when the exchange field is increased by an in-plane external
electric field.

\end{document}